\documentclass[preprint,aps,prd,showpacs,nofootinbib,superscriptaddress,floatfix]{revtex4-1}

\usepackage{amsmath}
\usepackage{slashed}
\usepackage{graphicx}
\usepackage{subfigure}
\usepackage{epstopdf}
\usepackage{epsfig}
\usepackage{amssymb}
\usepackage{bm}
\newcommand{\beq}{\begin{equation}}
\newcommand{\eeq}{\end{equation}}
\newcommand{\ov}{\overline}
\newcommand{\pa}{\partial}

\usepackage{color}

\begin{document}

\begin{flushright}
{\small EPHOU-14-018} 
\end{flushright}

\title{ Unification of SUSY breaking and GUT breaking}

\author{Tatsuo Kobayashi}
\affiliation{ Department of Physics, Hokkaido University, Sapporo 060-0810, Japan}

\author{Yuji Omura}
\affiliation{Department of Physics, Nagoya University, Nagoya 464-8602, Japan}

\date{\today}

\begin{abstract}
\noindent
We build explicit supersymmetric unification models
where grand unified gauge symmetry breaking and supersymmetry (SUSY) breaking are caused by the same sector.
Besides, the SM-charged particles are also predicted by the symmetry breaking sector, and they give the soft SUSY breaking terms through the so-called gauge mediation. 
We investigate the mass spectrums in an explicit model with $SU(5)$ and additional gauge groups,
and discuss its phenomenological aspects. Especially, nonzero A-term and B-term are generated at one-loop level
according to the mediation via the vector superfields, so that the electro-weak symmetry breaking and
$125$ GeV Higgs mass may be achieved by the large B-term and A-term even if the stop mass is around $1$ TeV.

\end{abstract}



\maketitle

\section{Introduction}

As well-known, the Standard Model (SM) is very successful in describing our nature,
and it is firmly established by the Higgs discovery at the LHC \cite{Higgs}.
There are still some ambiguities in not only the signal strength of the Higgs particle
but also the other observations such as flavor physics,
but it would be getting more difficult to consider new-physics effects in any signals. 

On the other hand, we are sure that the SM remains several mysteries about our nature:
the origin of the fermion generations, the hyper-charge assignment, the Higgs mass, and so on.
Many Beyond Standard Models (BSM) were proposed so far motivated by those mysteries, and some of them are expected to be found near future. One of the candidates is the supersymmetric grand unified theory (GUT),
which reveals the origin of the Higgs mass and the fermion charges.
There are some issues in Yukawa couplings, for instance, how to generate realistic Yukawa couplings
and heavy colored Higgs, but it succeeds in the charge quantization ($|Q_e+Q_p|<10^{-21}$ \cite{PDG})
and naturally deriving the electro-weak (EW) scale, if the supersymmetry (SUSY) scale $(\Lambda_{SUSY})$
is close to the EW scale. The supersymmetric GUT scenario is constrained by 
the observation of the proton decay, the direct search of SUSY particles, and the SM measurements.
Especially, the Higgs discovery around $125$ GeV may require high-scale SUSY $(\Lambda_{SUSY} \gg O(1){\rm TeV})$ \cite{Higgs1},
which may discard the strong motivation of SUSY, that is, the natural explanation of the EW scale.
Furthermore, the gauge coupling unification of supersymmetric $SU(5)$ GUT might be lost in high-scale SUSY,
depending on the mass spectrum of the SUSY particles.
The supersymmetric models could have so many parameters in the bottom-up approach, 
so that we could have some solutions for the Higgs mass and the gauge coupling unification.
However, it is very important to find how to derive such a specific SUSY mass spectrum.

In this paper, we propose an explicit supersymmetric GUT with $SU(5)_F \times SU(2) \times U(1)_{\phi} $ gauge
groups. We discard the miracle of the gauge coupling unification in the Minimal Supersymmetric SM (MSSM), but
SUSY breaking and GUT breaking sectors are unified. \footnote{This type of scenario has been proposed in Refs. \cite{Agashe:1998kg, Bajc:2006pa}.}
The SM-charged particles also appear after the symmetry breaking, so
the messenger fields for the gauge mediation is also introduced by the breaking sector in our model.
 \footnote{The messenger sector and SUSY breaking sector are unified, for instance, in Refs. \cite{directmediation}.}
The SM fields are only charged under the $SU(5)_F$ gauge group, so that
the charge quantization is realized. 

The breaking sector consists of
one $SU(5)_F$ adjoint plus singlet filed $(\Phi)$ and 
$SU(5)_F$ fundamental and anti-fundamental fields $(\phi,~\widetilde \phi)$.
The vector-like pairs $(\phi,~\widetilde \phi)$ are also charged under
$SU(2) \times U(1)_\phi$. As discussed in Ref. \cite{ISS}, this type of 
gauge theory causes SUSY breaking along with the gauge symmetry breaking.
In our model,  $SU(5)_F \times SU(2) \times U(1)_{\phi} $ symmetry
breaks down to the SM gauge groups, $SU(3)_c \times SU(2)_L \times U(1)_{Y} $,
where $SU(3)_c$ is from the subgroup of $SU(5)_F$, and $SU(2)_L \times U(1)_{Y} $
are the linear combinations of the subgroup of $SU(5)_F$ and $SU(2) \times U(1)_{\phi} $.
SUSY is broken by the F-component of the part of $\Phi$. 
After the symmetry breaking, SM-charged particles are generated by the fluctuation
of  $\Phi$ and $(\phi,~\widetilde \phi)$ around the vacuum expectation values (VEVs).
One interesting point is that the massive gauge boson of $SU(5)_F$ and the fermionic partners
could mediate the SUSY breaking effect through the gauge coupling with $\Phi$,
and play a crucial role in generating the non-zero A-term and B-term as discussed in Refs.\cite{Dermisek,Matos}.
It is well-known that SUSY-scale A-term could shift the upper bound on the lightest Higgs mass in 
the MSSM, even if squark is light,
 and the SUSY-scale B-term is required to realize the EW symmetry breaking.
Our A-term and B-term are given at one-loop level, so that 
they are the same order as the squark masses and gaugino masses. 
In fact, we will see that Higgs mass could be around $125$ GeV, even if $\Lambda_{SUSY}$
is less than $O(1)$ TeV, and the B-term could be consistent with the EW symmetry breaking.

In Sec. \ref{section2}, we introduce the SUSY and GUT breaking sector
in generic $ SU(N_F)_F \times SU(N)  \times U(1)_{\phi}$ gauge theory.
There, we discuss not only the symmetry breaking, but also
the behavior of the gauge couplings and soft SUSY breaking terms
according to the gauge mediation with the mediators of the chiral superfields
and the vector superfields. 
In Sec. \ref{section3}, we apply the breaking sector to the $ SU(5)_F \times SU(2)  \times U(1)_{\phi}$ gauge theory. 
As we mentioned above, an interesting aspect of this model is the improvement
of the consistency with the EW symmetry breaking and Higgs mass in the case with low-scale SUSY.
We investigate the soft SUSY breaking terms, and discuss how well it is achieved in our scenario.
In Sec. \ref{section4}, we give a comment on the possibility that
the breaking sector is applied to other GUT models.
Sec. \ref{section5} is devoted to the summary.
In Appendix \ref{appendix1}, we give the
mass spectrum in the SUSY breaking
sector. In Appendix  \ref{appendix2}, we show examples of mass spectrums in the MSSM sector.

\section{$SU(N) \times SU(N_F)_F \times U(1)_{\phi}$ gauge theory}
\label{section2}
In this section, we introduce the model which causes SUSY breaking together with gauge symmetry breaking,
based on Ref. \cite{ISS}.

We consider $SU(N_F)_F \times SU(N) \times  U(1)_{\phi}$ gauge theory with $N_F>N$.
The matter content is shown in Table \ref{table1}:
$\Phi$ is the $SU(N_F)_F$ adjoint plus singlet field and $(\phi, \widetilde \phi)$ pair
is the vector-like under $SU(N_F)_F \times SU(N) \times  U(1)_{\phi}$ gauge group.

\begin{table}[th]
\begin{center}
\begin{tabular}{c|ccc}
     & ~$\phi$~ & ~$\widetilde \phi$~ &$\Phi$      \\ \hline  
$SU(N_F)_F$  &  ${\bf N_F}$    & {\bf $\ov{{\bf N_F}}$}   &   {\bf adj$_{N_F}$}+{\bf 1}                  \\ 
$SU(N)$ & $ \ov{{\bf N}}$ &  ${\bf N} $     &  {\bf 1}   \\
$U(1)_{\phi}$ & $ Q_{\phi}$ &  $ -Q_{\phi}$    &  0 
    \end{tabular}
\caption{
\label{table1}%
{
Chiral superfields in $ SU(N_F)_F \times SU(N) \times U(1)_{\phi}$ gauge theory
}
}
\end{center}
\end{table}
The superpotential is given by 
\beq
W_R=- h Tr_N(  \widetilde \phi \Phi \phi) + h \Lambda_G  {\rm Tr}_{N_F}(\Phi),
\eeq
assigning $U(1)_R$ symmetry: the R-charge of $\Phi$ is $2$ and the R-charge of $(\phi, \widetilde \phi)$ is vanishing.
However, there would be an issue about how to break R-symmetry and how to avoid the
massless particle according the $U(1)_R$ symmetry breaking. Let us introduce explicit $U(1)_R$ breaking terms,
\beq
W_{\slashed{R}}= m_{\phi} Tr_N( \widetilde \phi \phi)+c,
\eeq
and discuss the superpotential as $W_{SB}=W_R +W_{\slashed{R}}$.
In Ref. \cite{ISS}, $W_R$ is generated, considering the dual side of $SU(N_F)_F \times SU(N+N_F)$
gauge theory with the $N_F$ vector-like pairs $(q_d, \widetilde q_d)$ of $SU(N+N_F)$ gauge group. $\Phi$
is interpreted as the composite operator as $ \Phi \equiv \widetilde q_d q_d$, and $h \Lambda_G  {\rm Tr}_{N_F}(\Phi)$
in $W_R$ corresponds to the mass term of the $(q_d,\widetilde q_d)$.

Some ideas to induce $W_{\slashed R}$ have been proposed in Ref. \cite{AKO1}, where the small wave-function factor
of $\Phi$ suppresses $\Phi^2$ and $\Phi^3$ terms according to the strong dynamics or the profile in the extra dimension. 
In Ref.\cite{AKO2}, the effect of the explicit R-symmetry breaking terms is well studied.
Here, we simply start the discussion from the superpotential $W_{SB}$ assuming
that such a mechanism, as discussed in Ref. \cite{AKO1}, works in underlying theories above the GUT, and study the symmetry breaking.  
In the global SUSY with canonical K\"ahler potential, 
the scalar potential is given by $V=|\pa_{\Phi} W_{SB}|^2+|\pa_{\phi} W_{SB}|^2+ |\pa_{\widetilde \phi}W_{SB}|^2$,
and SUSY vacua satisfy $\pa_{\Phi} W_{SB}=\pa_{\phi} W_{SB}=\pa_{\widetilde \phi} W_{SB}=0$.
In this model, $\pa_\Phi W_{SB}$ is given by
\beq
\pa_{\Phi_{ji}} W_{SB}= -h (  \phi  \widetilde \phi)_{ij} + h \Lambda_G \delta_{ij},
\eeq
and all elements cannot be vanishing, because $N_F \times N_F$ matrix $(\phi \widetilde \phi)$
has the rank $N$ $(<N_F)$.
This means that SUSY is broken by the F-components of $(N_F-N)$ elements in $\Phi$
and $SU(N_F)_F$ would be also broken.

Following Ref. \cite{ISS}, we decompose $\Phi$ and $(\phi, \tilde \phi)$ as 
\begin{eqnarray} 
\label{eq:decomposition}
\Phi &=& \left(
\begin{array}{cc}
\left( v_Y \right) {\bf 1}_N+ \Hat Y &\widetilde Z \\
 Z & \left( v_X \right) {\bf 1}_{\widetilde N}+ \Hat X
\end{array}
\right),  \\ \label{eq:decomposition2}
\phi &=& \left(
\begin{array}{c}
\left(v_{\chi} \right) {\bf 1}_N + \Hat \chi \\
\rho
\end{array}
\right), 
\widetilde \phi^T = \left(
\begin{array}{cc}
\left(v_{\chi}  \right) {\bf 1}_N + \Hat {\widetilde \chi}^T \\
\widetilde \rho^T
\end{array}
\right),
\end{eqnarray}
where $\Hat Y$, $\Hat \chi$ and $\Hat{\tilde \chi}$ are $N \times N$ matrices, 
$\Hat X$ is an $\widetilde N \times \widetilde N$ matrix ($\widetilde N =N_F-N$), 
$Z$ and $\rho$  ($\widetilde Z$ and $\widetilde \rho$) are 
$  N \times \widetilde N$ matrices ($  \widetilde N \times N$ matrices).
The VEVs, $v_{Y}$ and $v_\chi$, are fixed by the stationary conditions 
\begin{eqnarray}
v_Y&=& \frac{m_{\phi}}{h},  \\
v_\chi&=& \Lambda_G.  
\end{eqnarray} 
This solution also satisfies the D-flat conditions.
$v_X$ is a flat direction in global SUSY.
If we consider gravity and one-loop corrections, it would be stabilized at the nonzero value~\cite{ISS, AHKO}.

The nonzero VEVs break  $SU(N) \times SU(N_F)_F \times U(1)_{\phi}$ gauge symmetry to
$SU(\widetilde{N}) \times SU(N)_D \times  U(1)_Y$.
$SU(N)_D$ and $U(1)_Y$ are the linear combinations of the subgroups of $SU(N_F)_F$ and $SU(N) \times U(1)_{\phi}$.

\subsection{gauge bosons}
After the symmetry breaking, massive gauge bosons appear according to the Higgs mechanism.
Let us decompose the vector field $(V^{\mu}_F)$ for $SU(N_F)_F$ as
\begin{eqnarray}
V^{\mu}_F &=& \begin{pmatrix} W^{\mu}_F -a B'^{\mu}  && \frac{1}{\sqrt 2} (X^{\mu})^{\dagger}  \\ \frac{1}{\sqrt 2} X^{\mu}  && G^{\mu}  + \frac{N}{\widetilde N}a B'^{\mu} \end{pmatrix},  
\end{eqnarray}
where $a=\frac{\sqrt{\widetilde{N}}}{\sqrt{2N(N+\widetilde{N})}}$ is defined.
$W^{\mu}_F$ and $G^{\mu}$ are the adjoint representations of the subgroups of $SU(N_F)_F$: $SU(N)_F$ and $SU(\widetilde N)$. $X_\mu$ is the anti-fundamental and fundamental representations of $SU( N)_F \times SU(\widetilde N)$,
and $B'_{\mu}$ is the $U(1)_F$ vector field, where $U(1)_F$ is from $SU(N_F)_F$.
 
The nonzero VEVs generate the following mass terms,
\begin{eqnarray}
{\cal L}_g &=& M^2_X X^{\dagger}_{\mu} X^{\mu} + \frac{1}{2} M^2_{W'} W'^{A}_{\mu} W'^{A\mu} + \frac{1}{2} M^2_{Z'} Z'_{\mu} Z'^{\mu},  \\
M^2_X &=&g^2_F (v^2_{\chi} +\Delta v^2),   \\
M^2_{W'} &=&2 (g^2_F+g'^2_N) v^2_{\chi},   \\
M^2_{Z'} &=&4 N   (Q^2_{\phi} g^2_{\phi}+a^2g_F^2  ) v^2_{\chi},
\end{eqnarray}
where $\Delta v=v_X-v_Y$ is defined.
$W'^{A\mu}$ and $Z'^{\mu}$ are given by the linear combinations of $W^{A \mu}_F$ and $SU(N)$ gauge boson ($W^{A \mu}_N$), and  $B'^{\mu}$ and $U(1)_{\phi}$ gauge boson ($A^{\mu}_{\phi}$) respectively:
\begin{eqnarray}
\begin{pmatrix} B'^{\mu}  \\ A^{\mu}_{\phi} \end{pmatrix} &=& \begin{pmatrix} \cos \theta_Y && \sin \theta_{Y}
\\ -\sin \theta_{Y}  && \cos \theta_{Y} \end{pmatrix} \begin{pmatrix} B^{\mu}  \\ Z'^{\mu} \end{pmatrix},  \\
\begin{pmatrix}W^{A \mu}_F  \\ W^{A\mu}_{N} \end{pmatrix} &=& \begin{pmatrix} \cos \theta && -\sin \theta
\\ \sin \theta  && \cos \theta \end{pmatrix} \begin{pmatrix} W^{A\mu}  \\ W'^{A\mu} \end{pmatrix},
\end{eqnarray}
where $\cos \theta_Y$ and $\cos \theta$ are defined as
\begin{equation}
\cos \theta_Y =\frac{Q_{\phi}g_{\phi}}{\sqrt{Q_{\phi}^2g_{\phi}^2+a^2g_F^2}},~\cos \theta =\frac{g'_N}{\sqrt{g'^2_N+g_F^2}}.
\end{equation}
$G_{\mu}$, $W_{\mu}$, and $B_{\mu}$ are the gauge bosons for $SU(\widetilde{N}) \times SU(N)_D \times  U(1)_Y$ gauge symmetry, and their gauge couplings are given by 
\beq
g_N=g_F \cos \theta,~g_{\widetilde N}=g_F,~ g'_1=aNg_1=a N g_F \cos \theta_Y.
\eeq

\subsection{SM-charged fields from symmetry breaking sector}
According to the decomposition in Eqs. (\ref{eq:decomposition}) and (\ref{eq:decomposition2}),
we introduce the charge assignment of $(Z,\widetilde Z)$, $(\rho,\widetilde \rho)$, $Y$, $(\chi,\widetilde \chi)$, and $X$
in Table \ref{table2}. 
$Y$, $(\chi,\widetilde \chi)$, and  $X$ are the adjoint parts of $\Hat Y$, $(\Hat \chi, \Hat{\widetilde \chi})$, and $\Hat{X}$.
The singlet parts are not charged under the SM, and they are not so relevant to our analysis.
The mass matrices are studied in Appendix \ref{appendix1}.

\begin{table}[th]
\begin{center}
\begin{tabular}{c|cccc||ccc||c}
     &   ~ $Z$~ &  ~$\widetilde Z$ ~  & ~ $\rho$~ &  ~$\widetilde \rho$ ~  &  ~$Y$~ &~ $\chi$~ &  ~$\widetilde \chi$ ~   & ~$X$~     \\ \hline  
     $SU(\widetilde{N})$ & $\widetilde{ {\bf N}}$  & $\ov{\widetilde{ {\bf N}}}$ & $\widetilde{ {\bf N}}$  & $\ov{\widetilde{ {\bf N}}}$ &{\bf 1}  & {\bf 1}   & {\bf 1}  & {\bf adj$_{\widetilde{N}}$}       \\ 
$SU(N)_D$  &$ \ov{{\bf N}}$ & ${\bf N}$ & $\ov {\bf N}$ & $ {\bf N}$  &   {\bf adj$_N$}  &     {\bf adj$_N$}  &   {\bf adj$_N$}&  {\bf 1}                 \\ 
  $U(1)_Y$  &$ \frac{N + \widetilde N}{N \widetilde N} $ &  $\frac{-N - \widetilde N}{N  \widetilde N}$ &  $\frac{N+  \widetilde N}{N \widetilde N}$ &  $\frac{-N - \widetilde N}{N  \widetilde N}$  & 0 & 0 & 0 & 0
  \end{tabular}
\caption{
\label{table2}%
{
Extra Chiral superfields charged under the $SU(\widetilde{N}) \times SU(N)_D \times  U(1)_Y$. 
}
}
\end{center}
\end{table}

These fields obtain masses according to the nonzero VEVs, $v_{\chi}$, $v_Y$ and $v_X$ as we see in the Appendix \ref{appendix1}. They decouple at some scales above the EW scale.
In the next subsection, we investigate the RG flows of the gauge couplings including the threshold corrections
and discuss the soft SUSY breaking terms mediated by the heavy fields. 

\subsection{RG flows of the gauge couplings}
In this model, two kinds of symmetry breaking actually happen: one is $SU(N_F)_F$ breaking, $SU(N_F)_F \to SU(\widetilde{N}) \times SU(N)_F \times  U(1)_F$, and the other is $SU(N)_F \times U(1)_F \times SU(N) \times U(1)_{\phi}$ breaking: $SU(N)_F \times SU(N)  \to SU(N)_D $ and $ U(1)_F \times U(1)_{\phi} \to U(1)_Y$. The former is caused by $\Delta v$,
and the later is by $v_{\chi}$.
We consider a simple scenario assuming $\Delta v \gg v_{\chi}$.

As we see in Appendix \ref{appendix1}, there will be several intermediate scales, where
heavy particles in the symmetry breaking sector are decoupled and the RG flow of gauge couplings is modify.
According to the one-loop RG equations, the gauge couplings at the EW scale ($M_Z$) are evaluated as follows:
$SU(N)_F$, $SU(N)$ and $SU(N)_D$ gauge couplings $(\alpha_{F_N}, \alpha'_N, \alpha_N)$ are  
\begin{eqnarray} \label{gaugeN}
4 \pi \alpha^{-1}_{N}(M_Z)&=& 4 \pi \alpha^{-1}_{F_N}(T_{\chi_N})+ 4 \pi \alpha'^{-1}_{N}(T_{\chi_N})+ b_{N} \ln \left ( \frac{M_Z^2}{T^2_{\chi_N}} \right )+\Delta b^N_{\rm ex}  \left ( \frac{T_{\rm ex}^2}{\Lambda^2} \right ),  \\
 4 \pi \alpha^{-1}_{F_N}(T_{\chi_N})&=&4 \pi \alpha^{-1}_{G}(\Lambda)+\Delta b_N  \ln \left ( \frac{T_N^2}{\Lambda^2} \right )+(b_{N} -2N)  \ln \left ( \frac{T_{\chi_N}^2}{\Lambda^2} \right ),   \\
   4 \pi \alpha'^{-1}_{N}(T_{\chi_N})&=&4 \pi \alpha'^{-1}_{N}(\Lambda)+\Delta b_{\rho_N} 
  \ln \left ( \frac{T_{\rho_N}^2}{\Lambda^2} \right )-N \ln \left ( \frac{T_{\chi_N}^2}{\Lambda^2} \right ).
 \end{eqnarray}
$SU(\widetilde{N})$ gauge coupling $(\alpha_{\widetilde N})$ is   
\begin{eqnarray} \label{gaugeNt}
4 \pi \alpha^{-1}_{\widetilde N}(M_Z)&=&4 \pi \alpha^{-1}_{G}(\Lambda)+b_{\widetilde N}  \ln \left ( \frac{M_Z^2}{\Lambda^2} \right ) +\Delta b_{\widetilde N}  \ln \left ( \frac{T_{\widetilde N}^2}{\Lambda^2} \right )  \nonumber \\
&+&\Delta b^{\widetilde N}_{\rm ex}  \ln \left ( \frac{T_{\rm ex}^2}{\Lambda^2} \right )+\Delta b_{\rho_{\widetilde N}} 
  \ln \left ( \frac{T_{\rho_{\widetilde N}}^2}{\Lambda^2} \right )+\Delta b_{X}  \ln \left ( \frac{T_{\chi_{\widetilde N}}^2}{\Lambda^2} \right ).  \nonumber \\  
\end{eqnarray}    
 $U(1)_F$, $U(1)_{\phi}$, and $U(1)_Y$ gauge couplings $(\alpha_{F_1}, \alpha_\phi, \alpha_1)$  are
\begin{eqnarray} \label{gauge1}   
4 \pi \alpha^{-1}_{1}(M_Z)&=& 4 \pi \alpha^{-1}_{F_1}(T_{\chi_1})+ \frac{4 \pi a^2}{Q_{\phi}^2}  \alpha^{-1}_{\phi}(T_{\chi_1})+ b_{1} \ln \left ( \frac{M_Z^2}{T^2_{\chi_1}} \right )+\Delta b^1_{\rm ex}  \ln \left ( \frac{T_{ \rm ex}^2}{\Lambda^2} \right ),  \\
 4 \pi \alpha^{-1}_{F_1}(T_{\chi_1}) &=&4 \pi \alpha^{-1}_{G}(\Lambda)+\Delta b_1  \ln \left ( \frac{T_1^2}{\Lambda^2} \right )+\Delta b_{\rho_1} 
  \ln \left ( \frac{T_{\rho_1}^2}{\Lambda^2} \right )+(b_1 +\Delta b_{\chi_1})  \ln \left ( \frac{T_{\chi_1}^2}{\Lambda^2} \right ),   
  \\
   4 \pi \alpha^{-1}_{\phi}(T_{\chi_1}) &=&4 \pi \alpha^{-1}_{\phi}(\Lambda)+\Delta b_{\rho_{\phi}} 
  \ln \left ( \frac{T_{\rho_1}^2}{\Lambda^2} \right )+\Delta b_{\chi_{\phi}}  \ln \left ( \frac{T_{\chi_1}^2}{\Lambda^2} \right ).    
  \nonumber \\
\end{eqnarray}
$\Lambda$ is the cut-off scale and
$T_i$, $T_{\chi_i}$ and $T_{\rho_i}$ $(i=N,\widetilde{N},1)$ are the intermediate scales
where $X_{\mu}$, $\chi_i(\chi_{\widetilde N}\equiv X)$, and $\rho_i$ decouple respecitvely.  
According to the mass spectrums at each scale in Appendix \ref{appendix1},
$T_i$, $T_{\chi_i}$ and $T_{\rho_i}$ $(i=N,\widetilde{N},1)$ are estimated as
\begin{eqnarray}
(T_{N},T_{\rho_N},T_{\chi_N})&=&(M_X,h \Delta v, \sqrt{h}M_{G'}),  \\
(T_{\widetilde N},T_{\rho_{\widetilde N}},T_{\chi_{\widetilde N}})&=& (M_X,h \Delta v,m_X), \\
(T_{1},T_{\rho_1},T_{\chi_1})&=&   (M_X,h \Delta v, \sqrt{h} M_{Z'}).
\end{eqnarray}
The factor in front of each intermediate scale describes the freedom of the particles decoupling at the scale:
\begin{eqnarray}
(\Delta b_{N}, \Delta b_{\rho_N})&=&(2 (N_F-N),-\widetilde{N}),  \\
(\Delta b_{\widetilde{N}}, \Delta b_{\rho_{\widetilde{N}}},\Delta b_{X})&=&(2 (N_F-\widetilde{N}),-N,-\widetilde{N}),   \\
(\Delta b_{1},\Delta b_{\rho_{1}},\Delta b_{\chi_{1}})&=& \left( 2 N_F,-a^2N^2 \frac{2N}{\widetilde N} ,-2a^2N^2 \right ),  \\
(\Delta b_{\rho_{\phi}},\Delta b_{\chi_{\phi}})&=& (-2N \widetilde{N}Q^2_{\phi},-2N^2Q^2_{\phi}). 
\end{eqnarray}
We may also have to introduce additional particles charged under the gauge symmetry, in order to achieve realistic mass spectrums. For instance, colored Higgs would be necessary to derive the MSSM Higgs doublet at the low scale in Sec. \ref{section3},
and it is charged under $SU(\widetilde N) \times U(1)_F$ in our explicit model. 
Such an extra intermediate scale and the coefficient is defined as 
$T_{\rm ex}$ and $\Delta b^J_{\rm ex}$ $(J=\widetilde N , \, 1)$.

We also study the soft SUSY breaking terms of sfermions in the next subsection. Let us also introduce the wave function renormalization factor $( Z_q )$ for $SU(N)_F$-charged field $(q)$.
The one-loop renormalization group for $Z_q$ can be integrated analytically, if the Yukawa coupling is negligible, 
\begin{eqnarray}\label{wave-func}
\ln Z_q(M_Z) &=&\ln Z_q(\Lambda)+ \frac{2 c^q_G}{b_G} \ln \left ( \frac{\alpha_G(\Lambda)}{\alpha_G(T_i)} \right )+ \frac{2 c^q_i}{b_G-\Delta b_i} \ln \left ( \frac{\alpha_{F_i}(T_{i})}{\alpha_{F_i}(T_{\rm ex})} \right ) \nonumber \\
&+& \frac{2 c^q_i}{b_G-\Delta b_i-\Delta b^i_{\rm ex}} \ln \left ( \frac{\alpha_{F_i}(T_{\rm ex})}{\alpha_{F_i}(T_{\rho_i})} \right ) \nonumber \\
&+& \frac{2 c^q_i}{b_G-\Delta b_i-\widetilde{\Delta b_{\rho_i}}-\Delta b^i_{\rm ex}} \ln \left ( \frac{\alpha_{F_i}(T_{\rho_i})}{\alpha_{F_i}(T_{\chi_i})} \right ) + \frac{2 c^q_i}{b_i} \ln \left ( \frac{\alpha_{i}(T_{\chi_i})}{\alpha_{i}(M_Z)} \right ), 
\end{eqnarray} 
where $(\widetilde{\Delta b_{\rho_N}},\widetilde{\Delta b_{\rho_{\widetilde N}}},\widetilde{\Delta b_{\rho_1}})=(0,\Delta b_{\rho_{\widetilde N}},\Delta b_{\rho_1})$ is defined and $T_i \geq T_{\rm ex} \geq T_{\rho_i}$ is assumed.
$c^q_G$ and $c^q_i$ are the second Casimir of the field $q$, corresponding to the gauge groups.   
The masses squared of sfermions can be derived by the $v_X$-dependence in $Z_q$.
$v_X$ appears in the gauge couplings, so that $v_X$-dependence on the gauge couplings is only relevant
to the sfermion masses \cite{Giudice:1997ni}.

\subsection{Soft SUSY breaking terms}
\label{sec2-4}
Based on the above results, we investigate soft SUSY breaking terms which relate to particles 
charged under the gauge symmetry.
Soft SUSY breaking terms in $SU(\widetilde{N}) \times SU(N)_D \times  U(1)$ are calculated by
substituting $v_X+\theta^2F_X$ for $v_X$ in the gauge couplings \cite{Giudice:1997ni}.
Compared to typical gauge mediation, where messengers are only chiral superfields, massive gauge bosons and the fermionic partners also work as the mediators 
to generate the soft SUSY breaking terms, in our models \cite{Hisano,Intriligator:2010be,Dermisek,Matos}.

 In Eqs. (\ref{gaugeN}) , (\ref{gaugeNt}), and (\ref{gauge1}), the only intermediate scales, $T_i$, $T_{\rho_i}$, and $T_{\rm ex}$ depend on $v_X$.
This leads the masses $(M_{\widetilde N},M_N,M_1)$ of the gauginos,
which are the superpartner of $SU(N)_D \times SU(\widetilde{N}) \times U(1)$ gauge bosons, as follows:
\begin{eqnarray}
M_N (\mu)&=&- (\Delta b_N +\Delta b_{\rho_N}+\Delta b_{\rm ex}^{N} \xi_N) \frac{\alpha_{N} (\mu)}{4 \pi} \frac{F_X}{|\Delta v|}, \\
M_{\widetilde {N}} (\mu)&=&- (\Delta b_{\widetilde N} +\Delta b_{\rho_{\widetilde N}}+\Delta b_{\rm ex}^{\widetilde N} \xi_{\widetilde N}) \frac{\alpha_{{\widetilde N}} (\mu)}{4 \pi} \frac{F_X}{|\Delta v|}, \\
M_1(\mu)&=&- \left( \Delta b_{1} +\Delta b_{\rho_{1}}+\frac {a^2}{Q^2_{\phi}}\Delta b_{\rho_{\phi}}+\Delta b_{\rm ex}^{1} \xi_1 \right) \frac{\alpha_{1} (\mu)}{4 \pi} \frac{F_X}{|\Delta v|}.
\end{eqnarray}
$\xi_N$, $\xi_{\widetilde N}$, and $\xi_1$ describe the $v_X$ dependence on the mass scale of extra particles, $T_{\rm ex}$. For example, the holomorphic mass of extra particles may be given by $m_{\rm ex}+ \lambda_{\rm ex} (v_X+ \theta^2 F_X)$, where $m_{\rm ex}$ and $\lambda_{\rm ex}$ are a supersymmetric mass term and Yukawa coupling involving the extra particles. That is, the gaugino mass contribution of $ \ln (T_{\rm ex})$ would be proportional to $ \lambda_{\rm ex} F_X/m_{\rm ex}$, if $m_{\rm ex}$ is larger than $\lambda_{\rm ex} v_X$. In this case, $\xi_i$ is approximately given by 
$\xi_i=\lambda_{\rm ex} |\Delta v|/m_{\rm ex}$.

Let us consider the soft SUSY breaking terms corresponding to the trilinear (A-term)
and bilinear couplings (B-term) of the scalar components of the $SU(N_F)_F$-charged fields ($q_I$).
They are relevant to the $v_X$-dependence of the wave renormalization factor.
For instance, the A-terms corresponding to the Yukawa couplings $y_{IJK} q_I q_J q_K$ in the superpotential
are given by $A_{IJK}=A_I+A_J+A_K$, where $A_I =\frac{\pa \ln Z_I}{\pa \ln v_X} $ is defined and 
the trilinear coupling is described as $y_{IJK}A_{IJK} q_I q_J q_K$.

Eventually, $A_I$ is obtained from Eq. (\ref{wave-func}),
\begin{eqnarray}
A_I &=&\Biggl \{ 2 c^I_G\frac{\alpha_{G} (T_i)}{4 \pi}-\frac{2b_G c^I_i}{b_G-\Delta b_i} \frac{\alpha_{F_i} (T_i)}{4 \pi} \nonumber \\
&+&\left (\frac{2 c^I_i}{b_G-\Delta b_i}-\frac{2 c^I_i}{b_G-\Delta b_i -\Delta b^i_{\rm ex}} \right ) (b_G \xi+\Delta b_i (1-\xi)) \frac{\alpha_{F_i}(T_{\rm ex})}{4 \pi}  \nonumber \\
&+&\left (\frac{2 c^I_i}{b_G-\Delta b_i-\Delta b^i_{\rm ex}}-\frac{2 c^I_i}{b_G-\Delta b_i-\Delta b^i_{\rm ex} - \widetilde{\Delta b_{\rho_i}}} \right )(b_G-\Delta b^i_{\rm ex} (1- \xi)) \frac{\alpha_{F_i} (T_{\rho_i})}{4 \pi} \nonumber \\
&+& \left ( \frac{2(\Delta b_i+\Delta b^i_{\rm ex} \xi+\widetilde{\Delta b_{\rho_i} }) c^I_i }{b_G-\Delta b_i-\Delta b^i_{\rm ex}-\widetilde{\Delta b_{\rho_i}} } \right )
\frac{\alpha_{F_i} (T_{\chi_i})}{4 \pi}- \left ( \frac{2(\Delta b_i+\Delta b^i_{\rm ex} \xi+\Delta b'_{\rho_i} ) c^I_i }{b_i } \right )
\frac{\alpha_{i} (T_{\chi_i})}{4 \pi} \nonumber \\
&+&\left ( \frac{2(\Delta b_i+\Delta b^i_{\rm ex} \xi+\Delta b'_{\rho_i} ) c^I_i }{b_i } \right )
\frac{\alpha_{i} (\mu)}{4 \pi} \Biggl \} \frac{F_X}{|\Delta v|}, \label{A-term}
\end{eqnarray}
assuming $\xi_N=\xi_{\widetilde N}=\xi_1=\xi$. $\alpha_{F_{\widetilde N }} \equiv \alpha_{\widetilde N} $ is defined.

The masses squared ($m^2_q$) of $q$ could be also estimated by the Eq. (\ref{wave-func}),
seeing the $|v_X|^2$-dependence of $Z_q$ \cite{Giudice:1997ni}.
As discussed in Ref. \cite{Intriligator:2010be}, the gauge mediation with gauge messengers 
may contribute to the masses squared at the one-loop level,
if the gauge symmetry breaking and SUSY breaking are caused by the VEVs and F-components
of several fields. In our case, we simply assume $v_\chi \ll \Delta v$, so that
the gauge symmetry breaking and SUSY breaking are caused by only $\Delta v$ and
the F-component of $\Delta v$.\footnote{$\Delta v$ corresponds to the VEV of one adjoint field.} 
The one-loop correction is strongly suppressed by
$(v_{\chi}/\Delta v)^2$ according to Ref. \cite{Intriligator:2010be}, so that
we have to investigate the two-loop corrections, as discussed in Refs. \cite{Giudice:1997ni, Dermisek}.

Following Refs. \cite{Giudice:1997ni, Dermisek}, $m^2_q$ could be written as 
\begin{eqnarray}
m^2_q (\mu) &=&\Biggl \{ 2  b_G c^q_G\frac{\alpha^2_{G} (T_i)}{(4 \pi)^2}-\frac{2b^2_G c^q_i}{b_G-\Delta b_i} \frac{\alpha^2_{F_i} (T_i)}{(4 \pi)^2} \nonumber \\
&+&\left (\frac{2 c^q_i}{b_G-\Delta b_i}-\frac{2 c^q_i}{b_G-\Delta b_i -\Delta b^i_{\rm ex}} \right ) (b_G \xi+\Delta b_i (1-\xi))^2 \frac{\alpha^2_{F_i}(T_{\rm ex})}{(4 \pi)^2}  \nonumber \\
&+&\left (\frac{2 c^q_i}{b_G-\Delta b_i-\Delta b^i_{\rm ex}}-\frac{2 c^q_i}{b_G-\Delta b_i-\Delta b^i_{\rm ex} - \widetilde{\Delta b_{\rho_i}}} \right )(b_G-\Delta b^i_{\rm ex} (1- \xi))^2 \frac{\alpha^2_{F_i} (T_{\rho_i})}{(4 \pi)^2} \nonumber \\
&+& \left ( \frac{2(\Delta b_i+\Delta b^i_{\rm ex} \xi+\widetilde{\Delta b_{\rho_i} })^2 c^q_i }{b_G-\Delta b_i-\Delta b^i_{\rm ex}-\widetilde{\Delta b_{\rho_i}} } \right )
\frac{\alpha^2_{F_i} (T_{\chi_i})}{(4 \pi)^2}- \left ( \frac{2(\Delta b_i+\Delta b^i_{\rm ex} \xi+\Delta b'_{\rho_i} )^2 c^q_i }{b_i } \right )
\frac{\alpha^2_{i} (T_{\chi_i})}{(4 \pi)^2} \nonumber \\
&+&\left ( \frac{2(\Delta b_i+\Delta b^i_{\rm ex} \xi+\Delta b'_{\rho_i} )^2 c^q_i }{b_i } \right )
\frac{\alpha^2_{i} (\mu)}{(4 \pi)^2} \Biggl \} \frac{F_X^2}{|\Delta v|^2}, \label{squared mass}
\end{eqnarray}
where $\Delta b'_{\rho_i}$ is $(\Delta b'_{\rho_3},\Delta b'_{\rho_2},\Delta b'_{\rho_1})=(\Delta b_{\rho_3},\Delta b_{\rho_2},\Delta b_{\rho_1}+a^2\Delta b_{\rho_\phi}/{Q^2_{\phi}})$.

In the next section, we discuss one explicit model, where $SU(\widetilde{N}) \times SU(N)_D \times  U(1)_Y$
is the SM gauge groups corresponding to $(N_F,N)=(5,2)$.
In the explicit model, we see that a few parameters control all soft SUSY breaking terms
according to this analysis. Then, $\Lambda_{SUSY}$ is roughly given by $(\alpha_G/(4 \pi)) \times (F_X/|\Delta v|) $,
and A-term and B-term are of $O(\Lambda_{SUSY})$, which we could expect that are consistent with the condition for the EW symmetry breaking. We study the compatibility with the EW condition and the Higgs mass, in Sec. \ref{sec3-D}.

\section{$SU(5)_F \times SU(2) \times U(1)_{\phi}$ gauge theory: $(N_F,N)=(5,2)$}
\label{section3}
In this section, we consider a $SU(5)_F \times SU(2) \times U(1)_{\phi}$  gauge symmetric model,
which correspond to the $(N_F,N)=(5,2)$ case. We expect that the MSSM fields
are embedded in to ${\bf 10}$ and $\overline{{\bf 5}}$ representation as in the Georgi-Glashow $SU(5)$ GUT.
Involving ${\bf 5}$-representation Higgs $(H,\ov{H})$, the superpotential for the Yukawa couplings in the visible sector
is
\begin{equation}\label{visible superpotential}
W_{vis}=\Hat y^u_{kl} H {\bf 10}^k  {\bf 10}^l + \Hat y^d_{kl}  \ov{H} \ov{{\bf 5}}^k  {\bf 10}^l,  
\end{equation}
where $\ov{{\bf 5}}^k$ and ${\bf 10}^l$ are defined as the matter fields. As well-known, 
$\Hat y^u_{kl}$ and $\Hat y^d_{kl}$ may require $\Phi$ and $(\phi, \widetilde \phi)$ dependences in order to generate realistic mass matrices at the EW scale according to the higher-dimensional operators. Here, we simply assume that
the contributions to the soft SUSY breaking terms are enough small.

One serious problem in the $SU(5)$ GUT is how to generate the mass splitting between the colored Higgs and the
MSSM Higgs doublet. The mass of colored Higgs should be around the GUT scale to avoid the
too short life time of proton: $m_{H_c} \gtrsim 10^{16}$GeV$\times(1{\rm TeV}/\Lambda_{\rm SUSY})$ \cite{coloredHiggs}.
In our $SU(5)_F \times SU(2) \times U(1)_{\phi}$ model, the relevant terms to the Higgs masses is written as
\beq
W_{H}=\mu \ov{H}H + \lambda_H \ov{H} \Phi H.
\eeq
After the symmetry breaking, the colored Higgs mass and MSSM Higgs mass are given by $\mu+\lambda v_X$ and
$\mu+\lambda_H v_Y$. If $v_Y=m_{\phi}/h$ is the GUT scale, $\mu$ should be also around the GUT scale and then
the fine-tuning between $\mu$ and $\lambda v_Y$ is required: $\mu+\lambda_H v_Y \approx O(M_Z)$.
On the other hand, we could expect that the colored Higgs is enough heavy because of $\mu$, if there is no cancellation between $\mu$ and $\lambda_H v_X$.
Let us also consider the case that $v_X$ is the GUT scale. In this case, the MSSM Higgs mass could be light if $\mu$ and
$v_Y$ are around the weak scale, and the colored Higgs is heavy: $m_{H_c} \approx \lambda_H v_X$.

In both cases, the colored Higgs couples with $v_X+F_X \theta^2$, so it mediates the
SUSY breaking effect to the soft SUSY breaking terms.
The supersymmetric mass for $SU(2)_L$ Higgs doublet is $\mu_2 = \mu +\lambda_H v_Y \approx O(M_Z)$.
On the other hand, the colored-Higgs mass is $m_{H_c} = \mu+ \lambda_H v_X \approx \lambda_H (v_Y-v_X)$,
so that $\xi$ for the colored Higgs in soft SUSY breaking terms is approximately estimated as $\xi \approx sign(\lambda_H)$.  
The one-loop correction of $H_c$ to $m^2_q$ would be suppressed, because 
the $m_{H_c}$-dependence appears in $Z_q$ as $\ln (|m_{H_c}+\lambda_H F_X \theta^2|^2)$ according to the study in Ref. \cite{Giudice:1997ni}.
We could apply our analysis in Sec. \ref{sec2-4} to this scenario.

\subsection{gauge couplings}
In this model, $SU(5)_F \times SU(2) \times U(1)_{\phi}$ breaks down to 
the SM gauge group, $SU(3)_c \times SU(2)_L \times U(1)_Y$.
$SU(3)_c$ is the subgroup of $SU(5)_F$ and $SU(2)_L \times U(1)_Y$
are the linear combination of $SU(2)_F \times U(1)_F$ and $SU(2) \times U(1)_{\phi}$.

\begin{figure}[!t]
\begin{center}
{\epsfig{figure=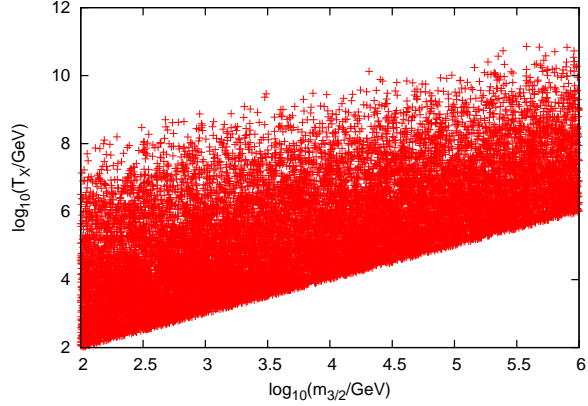,width=0.5\textwidth}}
\end{center}
\vspace{-0.5cm}
\caption{Gravitino mass ($m_{3/2}$) and the scale $T_X$ with $10^{16}$ GeV $\leq T_G \leq M_p$. $T_X$ should be small
to rase the GUT scale above $10^{16}$ GeV. $T_{\chi}$ is $T_{\chi} \approx \sqrt{M_p m_{3/2}}$.
The constraints, $T_{\rho}>T_{\chi}$ and $T_{X}>m_{3/2}$, are also assigned.
All gauge couplings and Yukawa couplings satisfy the perturbative bounds as $\alpha_{F_i} < 4 \pi.$
}
\label{fig1}
\end{figure}

\begin{figure}[!t]
\begin{center}
{\epsfig{figure=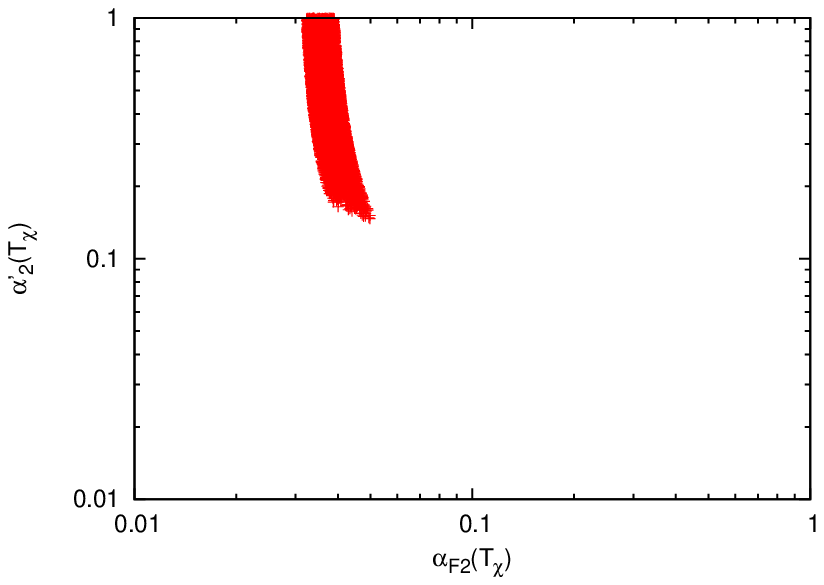,width=0.5\textwidth}}{\epsfig{figure=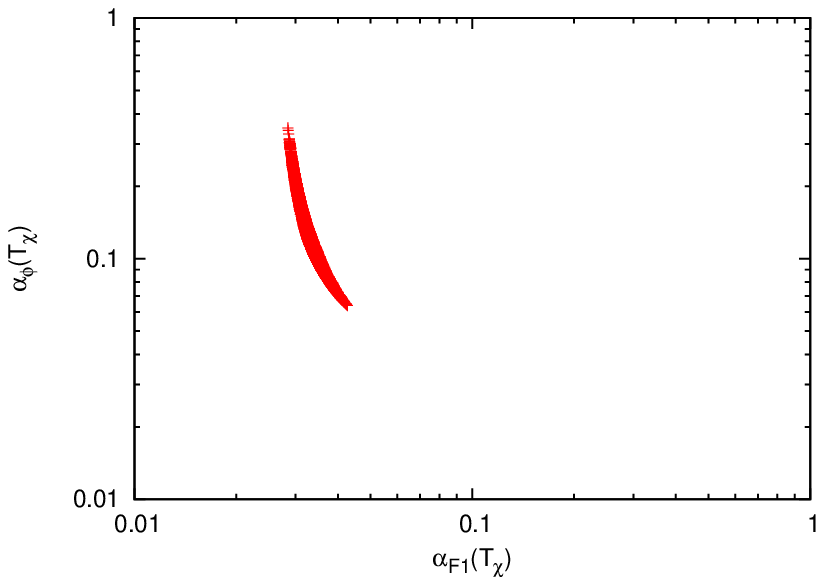,width=0.5\textwidth}}
\end{center}
\vspace{-0.5cm}
\caption{$\alpha_{F_2}$ vs. $\alpha'_2$ and $\alpha_{F_1}$ vs. $\alpha_\phi$ at the symmetry breaking scale, $T_{\chi}$.
}
\label{fig2}
\end{figure}

\begin{figure}[!t]
\begin{center}
{\epsfig{figure=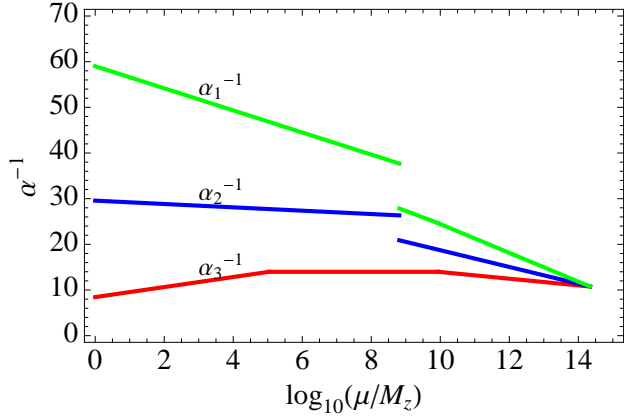,width=0.5\textwidth}}
\end{center}
\vspace{-0.5cm}
\caption{RG flows of the subgroups of $SU(5)_F$, 
with $T_X=10^7$ GeV, $T_\chi=3.8 \times 10^{10}$ GeV, $T_\rho=7.9 \times 10^{11}$ GeV, and $T_{GUT}=2 \times 10^{16}$ GeV. The green, blue, and red lines correspond to the gauge couplings 
of $U(1)_Y \times SU(2)_L \times SU(3)_c$ below $T_\chi$ and  $U(1)_F \times SU(2)_F \times SU(3)_c$ above $T_\chi$ respectively. The input parameters for the couplings are in Eq. (\ref{gauge-PDG}).
}
\label{fig0}
\end{figure}

On the other hand, there are several intermediate scales: $(T_G,T_\rho,T_\chi,T_X)$.
\footnote{$T_{\chi_1} =T_{\chi_2}$, $T_{\rho_1}=T_{\rho_3}$, and $T_G=T_1=T_2=T_3$ are assumed.}
$T_G$ is the GUT scale, where $X_{\mu}$ decouples, and $T_\rho$ is 
the messenger scale fixed by the parameter $h$ and the GUT scale.
$T_{\chi}$ is interpreted as the SUSY breaking scale, because $T_{\chi} \approx \sqrt{F_X}=\sqrt{M_p m_{3/2}}$,
so that it is almost fixed around $O(10^{10})$ GeV when $m_{3/2}=O(100)$ GeV.
$T_X$ is fixed by the mass scale of $X$ ($m_X$), which is massless at the tree-level.
$X$ could be expected to be $O(\Lambda_{SUSY})$, because the one-loop corrections shift the mass,
but it may be difficult to clearly fix the masses of bosonic and fermonic $X$ in our model.
Let us simply treat $m_X$ as the free parameter, and
Fig. \ref{fig1} shows the allowed region for $T_X$, which may not be far from $O(\Lambda_{SUSY})$. 
Fig. \ref{fig2} shows the gauge couplings, $(\alpha_{F2},\alpha_{2}, \alpha_{F1},\alpha_{\phi})$
at the SUSY breaking scale.
Fig. \ref{fig0} describes RG flows of the gauge couplings $(\alpha_3, \alpha_{2} (\alpha_{F_2}), \alpha_1(\alpha_{F_1}))$,
when $T_X=10^7$ GeV, $T_\chi=3.8 \times 10^{10}$ GeV, $T_\rho=7.9 \times 10^{11}$ GeV, and $T_{GUT}=2 \times 10^{16}$ GeV.

\subsection{soft SUSY breaking terms}
We qualitatively evaluate the soft SUSY breaking terms in this scenario.
According to the analysis in Sec. \ref{sec2-4}, the gaugino masses at $\mu < T_{\chi}$ are written as
\begin{eqnarray}
M_2 (\mu)&=&- 3 \frac{\alpha_{2} (\mu)}{4 \pi} \frac{F_X}{|\Delta v|}, \\ \label{gluino}
M_{3} (\mu)&=&- (2-\xi) \frac{\alpha_{{3}} (\mu)}{4 \pi} \frac{F_X}{|\Delta v|}, \\ 
M_1(\mu)&=&- \left( \frac{37-2 \xi }{5} \right) \frac{\alpha_{1} (\mu)}{4 \pi} \frac{F_X}{|\Delta v|}.
\end{eqnarray}
Let us consider the case with $\xi=1$ and the gaugino masses at the EW scale.
The gauge couplings at the EW scale are \cite{PDG}
\beq\label{gauge-PDG}
\alpha_1(M_Z) \approx 0.01695, ~\alpha_2(M_Z) \approx 0.03382, ~\alpha_3(M_Z) \approx 0.1185,
\eeq
so that we could derive the following mass relation:
\beq
  \frac{M_1(M_Z)}{M_3(M_Z)} \approx 1.001,~  \frac{M_2(M_Z)}{M_3(M_Z)}  \approx 0.856.
\eeq
The masses are almost degenerate, and this may be a specific feature of the gauge messenger model \cite{Dermisek,Bae}.\footnote{
The gaugino masses are degenerate in the TeV-scale mirage mediation scenario, too \cite{Choi:2005hd}.}
 If all intermediate scales are close to the GUT scale, the fine-tuning of $\mu$ term may be drastically reduced,
as discussed in Ref. \cite{AKO3}. Fig. \ref{fig2} tells us that the extra $SU(3)$-adjoint field reside in the low-scale,
so that the condition for the small $\mu$-term would be modified.
The one-loop running correction of $m^2_{H_u}$ with  $T_X=10^{7}$ GeV from $T_\chi$ to $M_Z$ is estimated as 
\beq
\Delta m^2_{H_u} \approx -0.276 M_3(M_Z)^2-0.047 M_2(M_Z) M_3(M_Z) +0.221 M_2(M_Z)^2 + \dots,
\eeq
where the ellipsis denotes the terms including A-term and scalar masses and those are not important 
when they are comparable to the gluino mass.
This leads that the condition to cancel the large contribution of gluino is $M_2/M_3(M_Z) \approx 1.23$,
which suggests the almost degenerate mass spectrum.
However, we have a large A-term contribution to $\Delta m^2_{H_u}$ in our model, so that
it may be difficult to avoid a certain fine-tuning even if the gaugino masses are degenerate.

According to Eqs. (\ref{squared mass}) and (\ref{A-term}),
the masses squared of superpartners and A-term are evaluated explicitly. 
Setting $T_G  = T_{H_c} > T_{\rho}>T_{\chi}$ and $\xi=1$, 
stop masses at $T_\chi$ are given by
\begin{eqnarray}
m_Q^2(T_{\chi}) &\approx& \left (8.83 -6.67 \frac{\alpha^2_{3}(T_{\rho})}{\alpha^2_G}-10.80 \frac{\alpha^2_{F_2}(T_{\chi})}{\alpha^2_G} -0.33 \frac{\alpha^2_{F_1}(T_{\chi})}{\alpha^2_G} \right )\Lambda^2_{SUSY}, \\
m_U^2(T_{\chi}) &\approx& \left (8.60 -6.67 \frac{\alpha^2_{3}(T_{\rho})}{\alpha^2_G}-5.30 \frac{\alpha^2_{F_1}(T_{\chi})}{\alpha^2_G}  -0.01 \frac{\alpha^2_{F_1}(T_{\rho})}{\alpha^2_G} \right )\Lambda^2_{SUSY}. 
\end{eqnarray}
As we see, large stop masses are generated by the large second casimir $(c^t_2=18/5)$, but 
they might be driven to the tachyonic if $T_{\chi}$ and $T_{\rho}$ are close to the GUT scale.
The SUSY scale $(\Lambda_{SUSY})$ from the gauge mediation is defined as 
\beq
\Lambda_{SUSY} = \frac{\alpha_G}{ (4 \pi)}  \frac{F_X}{|\Delta v|} \approx \frac{\alpha_G}{ (4 \pi)}  \frac{M_p}{T_{G}} m_{3/2}.
\eeq
$\alpha_G$ is of $O(0.1)$ when $T_G$ is around $10^{16}$ GeV, so that
$\Lambda_{SUSY}$ might be compatible with $m_{3/2}$.
If $T_G$ is smaller, the situation, $\Lambda_{SUSY} \gg m_{3/2}$, is achieved but suffers from the constraint
from proton decay. The correction from the gravity mediation is naively estimated as $O(m_{3/2})$.
It is almost the same order as the one from the gauge mediation in our model, and it may make it difficult to
control flavors. In fact, the gauge-mediation contributions are typically at least $5$ times as large as the gravitino mass in our model, as we see in Table \ref{table4}. In this case, we could expect the gravity-mediation effect is sub-dominant,
and the SUSY scale is governed by the gauge-mediation. 
However, the gravity-mediation contribution should be $O(10^{-2})$ times suppressed, if it contributes to the
sparticles masses squared flavor-universally \cite{Gabbiani:1996hi}.      
In order to realize such a suppression and control flavor in the MSSM, we have to consider flavor symmetry or some dynamics above the GUT scale, as discussed in Refs. \cite{conformal}. \footnote{In fact, such strong dynamics 
has been proposed not only to suppress flavor changing currents but also to realize the superpotential $W_{SB}$ in Sec. \ref{section2} \cite{AKO1}.  }  Indeed, explicit contributions on soft masses
through the gravity mediation depend on the UV completion of our model.
In this letter, one of our main motivations is to achieve $125$ GeV Higgs mass and realistic EW symmetry breaking,
which may be independent of this issue about the constraint from flavor physics,
so that we will discuss our SUSY mass spectrums assuming that the gauge-mediation is dominant. 
The underlying theory above the GUT scale will be studied in Ref. \cite{progress}.

$A_t$, which is the trilinear coupling of stops ($\widetilde t$) as $y_t A_t \widetilde{ t}_L H_u \widetilde{ t}_R$ is given by 
\beq
A_t(T_{\chi}) \approx \left (22.57 -8.00 \frac{\alpha_3 (T_{\rho})}{\alpha_G} -3.6 \frac{\alpha_{F_2} (T_{\chi})}{\alpha_G} -0.98 \frac{\alpha_{F_1} (T_{\chi})}{\alpha_G} +0.01 \frac{\alpha_{F_1} (T_{\rho})}{\alpha_G}  \right ) \Lambda_{SUSY},
\eeq
and the B-term, which is the bilinear coupling of two Higgs $\mu B H_u H_d$, is estimated as
\beq
B(T_{\chi}) \approx \left (10.27 -3.60 \frac{\alpha_{F_2} (T_{\chi})}{\alpha_G} -0.68 \frac{\alpha_{F_1} (T_{\chi})}{\alpha_G}+ 0.01 \frac{\alpha_{F_1} (T_{\rho})}{\alpha_G} \right ) \Lambda_{SUSY}.
\eeq
As we see, the A-term and B-term might be large as $O(10) \Lambda_{SUSY}$.
This may be good to achieve the EW symmetry breaking, but too large A-term makes the stop masses tachyonic 
because of the running correction such as
\beq
\Delta m^2_U(M_Z) \approx-0.08 A_t(T_{\chi})^2+1.54 M_3(T_{\chi})^2-0.15A_t(T_{\chi})M_3(T_{\chi}).
\eeq
In our model, the gluino mass $M_3$ is relatively small as wee see in Eq. (\ref{gluino}),
so $\Delta m^2_U(M_Z)$ becomes easily negative and stop mass becomes tachyonic even if 
the positive $m^2_U$ is generated at the SUSY breaking scale $T_{\chi}$.
In order to avoid the tachyonic stop masses, we add an extra contribution to the gluino mass, as we see below.

\subsection{Shift of the gluino mass}
We consider an extra term, which contributes to the gluino mass,
\beq
W= \frac{1}{\Lambda_0} Tr_5 (\Phi W_{5} W_{5}).
\eeq
There are several ways to introduce this term, such as gravity effect.
Here, we simply assume that $N_{ \rm extra}$ extra heavy $SU(5)$ vector-like pairs ($\psi, \ov{\psi}$) with the masses
$ \ov{\psi} (\Lambda_0 + \lambda_X \Phi) \psi$ induce this term, integrating out them at the scale $\Lambda_0$.
After the $SU(5)$ breaking, the gauge coupling would have the extra $v_X$ dependence as 
\beq
\alpha^{-1}_3 \to \alpha^{-1}_3 - \frac{N_{\rm extra}}{4 \pi}  \ln \left ( \frac{(| \Lambda_0+ \lambda_X (v_X+ F_X \theta^2) |^2)}{\Lambda^2} \right ).
\eeq
 This additional coupling could shift the gluino mass as
\beq
M_3 \to M_3- \frac{\alpha_{3}N_{eff}}{4 \pi}  \frac{F_X}{|\Delta v|},
\eeq
where $N_{eff}$ may not be $N_{ \rm extra}$ because of the scale difference between $\Lambda_0$ and the GUT scale.
Including $N_{eff}$, the gluino mass becomes
\beq
M_{3} (\mu)=- (1-N_{eff}) \frac{\alpha_{{3}} (\mu)}{4 \pi} \frac{F_X}{|\Delta v|},
\eeq
so $N_{eff}$ should be bigger than $2$ in order to shift $M_3$.
In fact, we discuss large $N_{eff}$ cases and find that $N_{eff}$ enables us to evade the negative squared masses
and achieve the large SM Higgs mass.
\subsection{Consistency with the Higgs mass and the EW symmetry breaking}
\label{sec3-D}
\begin{figure}[!t]
\begin{center}
{\epsfig{figure=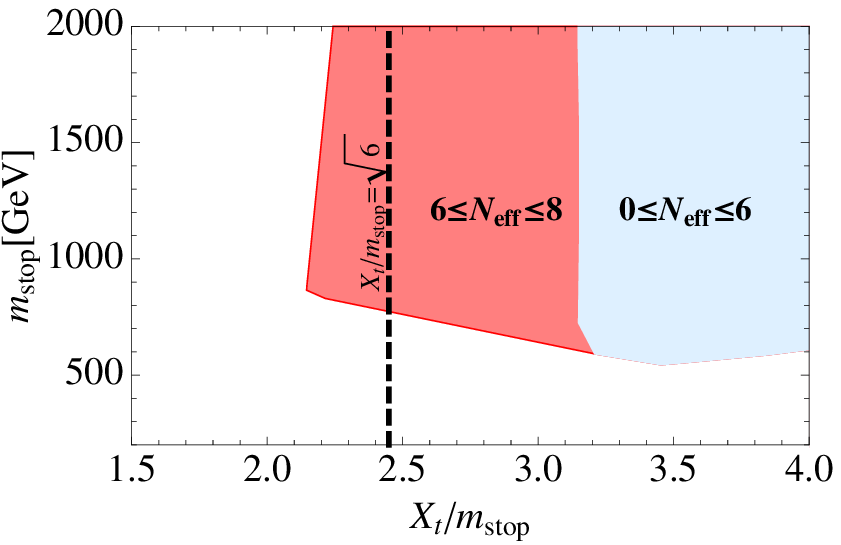,width=0.5\textwidth}}{\epsfig{figure=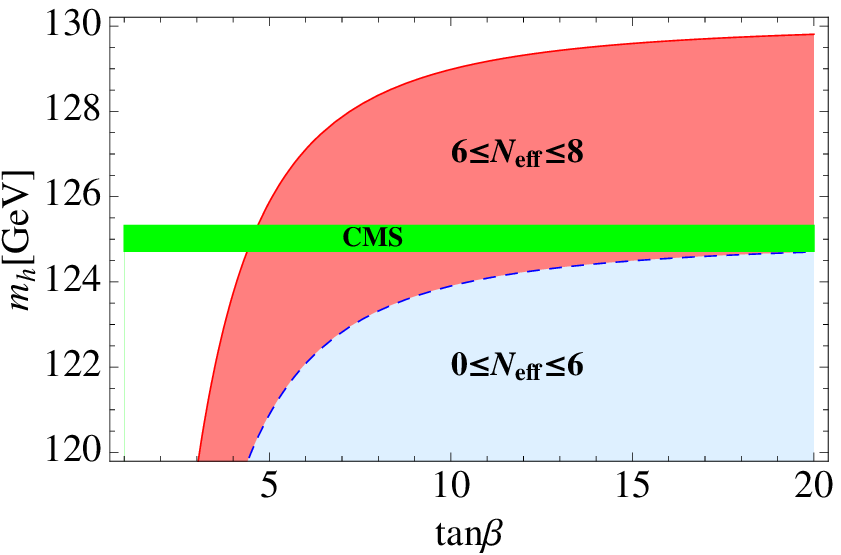,width=0.5\textwidth}}
\end{center}
\vspace{-0.5cm}
\caption{$X_t/m_{stop}$ vs. $m_{stop}$ and $\tan \beta$ vs. the lightest Higgs mass in the case with $(T_{\rm GUT},T_X)=(2 \times 10^{16} {\rm GeV},$ $10^{7} {\rm GeV})$ and $0 \leq N_{eff} \leq 6$ (light blue), $6 \leq N_{eff} \leq 8$ (light red). The dashed line corresponds to $X_t/m_{stop}=\sqrt{6}$. In the right figure, $m_h$ is calculated at the two-loop level using $m_t=172.9$ GeV, and $m_{stop}$ is lighter than $2$ TeV. The green band is the CMS result on Higgs mass from $h \to \gamma \gamma$, $ZZ$ channels \cite{Higgs-CMS}.
}
\label{fig3}
\end{figure}

One issue in supersymmetric models is how to realize the $\mu$ and $B$ terms
which are consistent with the EW scale. Especially, $\mu$ relates to the lightest Higgs mass,
because of the upper bound in MSSM, so that the recent Higgs discovery with the mass 
$125$ GeV may impose unnatural SUSY scenarios on us.
In fact, $125$ GeV Higgs mass may require $\Lambda_{SUSY} \gtrsim O(10)$ TeV in the simple scenarios as
discussed in Ref. \cite{Higgs1}. $O(10)$-TeV SUSY scale would require $0.01 \%$ fine-tuning against $\mu$
without any cancellation in $m^2_{H_u}$.
As pointed out in Refs. \cite{Higgs2,Higgs3}, it is known that a special relation between $A_t$ and
squark mass relaxes the fine-tuning, maximizing the loop corrections in the Higgs mass in the MSSM.
This relation is so-called ``maximal mixing" and described as $X_t/ m_{stop} = \sqrt{6}$, where
$X_t=A_t- \mu/\tan \beta$ and $m_{stop}^2=\sqrt{m_{Q}^2 m_{U}^2}$ are defined. 
If this relation is satisfied, the $125$ GeV Higgs mass could be achieved even if the stop is light.
We can see our prediction on $X_t$ and the upper bound on the Higgs mass in the case with $0 \leq N_{eff} \leq 6$ (light blue), $6 \leq N_{eff} \leq 8$ (light red) in Fig. \ref{fig3}.
On the all regions, all masses squared of the superpartners are positive
and $(T_G,T_X)$ are fixed at $(2 \times 10^{16} {\rm GeV},$ $10^{7} {\rm GeV})$.
We find that our A-term is too large to realize $X_t/ m_{stop} = \sqrt{6}$, but
the maximal mixing could be achieved, if we allow large $N_{eff}$,
and enhance the Higgs mass, even if $m_{stop}$ is around $1$ TeV.

\begin{figure}[!t]
\begin{center}
{\epsfig{figure=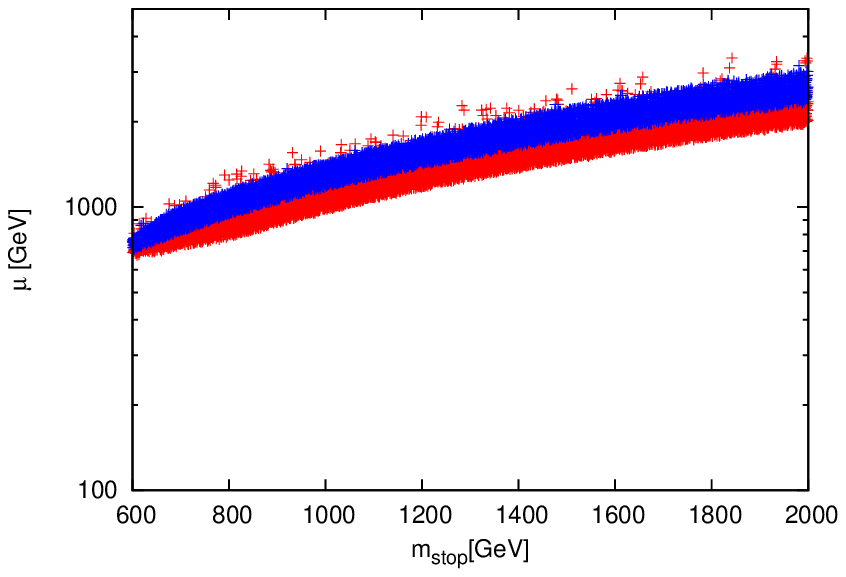,width=0.5\textwidth}}{\epsfig{figure=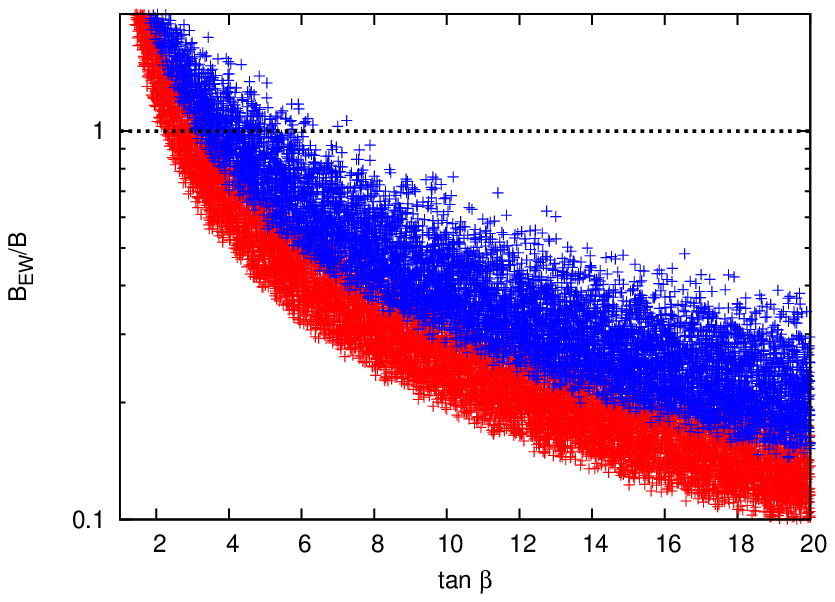,width=0.5\textwidth}}
\end{center}
\vspace{-0.5cm}
\caption{$m_{stop}$ vs. $\mu$ and $ \tan \beta$ vs. B-term in the case with $(T_{\rm GUT},T_X)=(2 \times 10^{16} {\rm GeV},$ $10^{7} {\rm GeV})$ and $0 \leq N_{eff} \leq$ (blue), $6 \leq N_{eff} \leq 8$ (red). In the right figure, $m_{stop}$ is lighter than $2$ TeV.
The dashed line is consistent with the condition for the EW symmetry breaking.
}
\label{fig4}
\end{figure}

On the other hand, we notice that there is no special cancellation in $m^2_{H_u}$ and $m^2_{H_d}$,
as we see in Fig. \ref{fig4}. Large $m_{stop}$ corresponds to large $\mu$, so that $1$-TeV squark mass
requires $1 \%$ fine-tuning against $\mu$.
The right figure in Fig. \ref{fig4} shows that small $\tan \beta$ is consistent with the EW symmetry breaking.
$B_{\rm EW}$ is the value to realize the EW symmetry breaking,
\beq
B_{\rm EW} \equiv - \frac{1}{2\mu} \{ (m^2_{H_d}-m^2_{H_u}) \tan 2 \beta +M_Z^2 \sin 2 \beta \},
\eeq
and $B$ is our prediction via the gauge mediation. 
It seems that $2 \lesssim \tan \beta \lesssim 6$ is necessary 
to achieve $125$ GeV Higgs mass. 
The $\tan \beta$ region may be inconsistent with the one required by $125$ GeV Higgs ($\tan \beta \gtrsim 4$)
with $m_{stop} \leq 2$ TeV.
Table \ref{table4} in Appendix \ref{appendix2} shows the parameter sets in our model, which satisfy $m_h \approx 125$ GeV and $|B_{\rm EW}/B| \approx 1$. There, $m_{stop}$ and $|\mu|$ are around $3$ TeV, and $O(0.1)$ \% fine-tuning
is required against $\mu$ term.

\section{$SU(5)_F \times SU(3) \times U(1)_{\phi}$ gauge theory: $(N_F,N)=(5,3)$}
\label{section4}
Our symmetry breaking model could be embed into other type GUT model.
One simple example would be the $SU(5)_F \times SU(3) \times U(1)_{\phi}$ gauge symmetric model, and
we could consider the same setup as in the $SU(5)_F \times SU(2) \times U(1)_{\phi}$ gauge theory.
The visible sector is given by Eq. (\ref{visible superpotential}).
However, the modification of the Higgs sector may be required because $\lambda \ov{H} \Phi H$ term gives the
very large B-term, $\lambda F_X H_u H_d$. There may be a solution to realize the EW symmetry breaking,
but the serious fine-tuning may be required. 
Here, we consider another solution to shift the colored Higgs mass which maybe favor high-scale SUSY.

We introduce $SU(3)$ vector-like fields $(H_3, \ov{H}_3)$ and assign $Z_3$ symmetry to the fields as in Table \ref{table3}.
$Z_3$ symmetry is broken by the VEV of $S$.
The superpotential for the Higgs sector is given by
\begin{equation}\label{eq:superpotential-Higgs}
W_{H}= \lambda_S  S \ov{H}H+ \lambda_{\phi}  \ov{H} \ov{H}_3 \phi+ \lambda_{ \widetilde \phi}  \widetilde{\phi} H_3 H +\frac{ \lambda_3}{3}  S^3.   
\end{equation}

\begin{table}[th]
\begin{center}
\begin{tabular}{c|cccc||ccc||ccc}
     &   ~$\ov {\bf 5}_i$~ &  ~${\bf 10}_i$~  & ~$H$~     & ~$\ov{H}$~  & ~$H_3$~ & ~$\ov{H}_3$~ & ~$S$~   & ~$\phi$~ & ~$\widetilde \phi$~ &$\Phi$      \\ \hline  
$SU(5)_F$  &$\ov {\bf 5}$ &  {\bf 10} &  {\bf 5} &  $\ov{{\bf 5}}$ & {\bf 1}   &   {\bf 1}   & {\bf 1}  &  {\bf 5}    & {\bf $\ov{{\bf 5}}$}   &   {\bf adj$_5$}+{\bf 1}                  \\ 
$SU(3)$ & {\bf 1}  & {\bf 1} &{\bf 1} &{\bf 1} &{\bf 3}  & $\ov {\bf 3} $  & {\bf 1}  & {\bf 3} &  $\ov {\bf 3} $     &  {\bf 1}       \\ 
  $U(1)_{\phi}$  & 0  & 0 & 0 & 0  & $Q_{\phi}$ & $-Q_{\phi}$ & 0 & $Q_{\phi}$ &  $-Q_{\phi}$ &  0    \\ \hline 
$Z_3$&  ${\bf \omega}$ & ${\bf \omega}$& ${\bf \omega}$& ${\bf \omega}$& ${\bf \omega}^2$& ${\bf \omega}^2$& ${\bf \omega}$& {\bf 1}& {\bf 1}& {\bf 1}
\end{tabular}
\caption{
\label{table3}%
{
Chiral superfields in $SU(5)_F \times SU(3) \times U(1)$ gauge theory
}
}
\end{center}
\end{table}

After the GUT symmetry breaking, $(H,\ov{H})$ are decomposed as
 $((H_u,H'_3),(H_d,\ov{H}'_3))$ and the mass terms for 
$(H_3, \ov{H}'_3)$ and $(H'_3, \ov{H}_3)$ pairs appear as
\beq
W^{eff}_{ H}= \lambda_{\phi} v_{\chi} \ov{H}'_3  \ov{H}_3 + \lambda_{ \widetilde \phi}  v_{\chi}  H_3 H'_3. 
\eeq
$H_u$ and $H_d$ correspond to the Higgs $SU(2)_L$ doublets in MSSM, and they 
could get the supersymmetric mass term according to the nonzero VEV of $S$. 
In Refs. \cite{productGUT}, we can see 
not only the $SU(5)_F \times SU(2) \times U(1)_{\phi}$-type but also this type of product-GUT.

In order to avoid the bound from the proton decay caused by the five dimensional operators, 
$v_{\chi}$ should be large as
\beq
 v_{\chi} \gtrsim 10^{16} {\rm GeV}  \times \left( \frac{1 {\rm TeV}}{ \Lambda_{\rm SUSY}} \right ).
\eeq
$F_X$ is given by $-h v^2_{\chi}$, so that very tiny $h$ is necessary to achieve the low-scale SUSY.
When $v_{\chi} \approx 10^{16}$ GeV and $\Lambda_{\rm SUSY}=1$ TeV are set,
$h$ should be around $O(10^{-10})$, because of
\beq
h =  \frac{4 \pi}{\alpha_G} \frac{\Delta v}{v^2_{\chi}} \Lambda_{\rm SUSY} \approx 10^{-10} \times \left ( \frac{10^{16} \rm GeV}{v_{\chi}} \right )^2  \left ( \frac{ \Lambda_{\rm SUSY}}{1 \rm TeV} \right ) \lesssim 10^{-10} \times \left ( \frac{\Lambda_{\rm SUSY}}{1 \rm TeV} \right )^3.
\eeq
We conclude that high-scale SUSY is favored to avoid such an extremely small $h$.

We can consider the applications of our symmetry breaking models to the other BSMs, such as 
\begin{itemize}
\item $SU(3)_c \times SU(2)_L \times SU(2)_R \times U(1)_{B-L} \rightarrow  SU(3)_c \times SU(2)_L \times U(1)_Y,$
\item $SU(4) \times SU(2)_L \times U(1) \rightarrow  SU(3)_c \times SU(2)_L \times U(1)_Y.$
\end{itemize}
We would study such patterns elsewhere \cite{progress}.
In these models, all of chiral superfields appear as adjoint representations and 
bi-fundamental representations.
Such models can be constructed in D-brane models, e.g. intersecting/magnetized D-brane models
(see for a review \cite{Blumenhagen:2006ci,Ibanez} and references therein).
Thus, the above models are interesting from the viewpoint of superstring theory.

\section{Summary}
\label{section5}
The MSSM is one of the attractive BSMs to solve the hierarchy problem in the SM
and it may be expected to be found near future.
One big issue in the MSSM is how to control the SUSY breaking parameters,
so that many ideas and works on spontaneous SUSY breaking and 
mediation mechanisms of the SUSY breaking effects have been discussed so far.
In this paper, we proposed an explicit and simple supersymmetric model,
where the spontaneous SUSY breaking and GUT breaking are achieved
by the same sector. The origin of the hyper-charge assignment in the MSSM 
is also explained by the analogy with the Georgi-Glashow $SU(5)$ GUT.
The SM-charged particles are also introduced by the
breaking sector, so that we could also predict the soft SUSY breaking terms
via the gauge mediation with the gauge and chiral messenger superfields.
The crucial role of the gauge-messenger mediation is to induce large A-terms and
B-terms at the one-loop level. We investigated the scenario with light superpartners that
such a large A-term realizes the maximal mixing and shift the lightest Higgs mass.
In fact, we have to introduce additional contribution to the gluino mass,
but $125$ GeV Higgs mass could be achieved, even if stop is light.
$m_{stop}$ should be as light as possible to relax the fine-tuning of $\mu$ parameter.
On the other hand, the one-loop B-term could be also consistent with the EW symmetry breaking,
if $\tan \beta$ is within $2 \lesssim \tan \beta \lesssim 6$. Such small $\tan \beta$ may require large stop mass,
as we see in Figs. \ref{fig3} and \ref{fig4}.
In fact, we see that about $3$ TeV $m_{stop}$ can achieve $125$ GeV Higgs mass and the EW symmetry breaking
in Table \ref{table4}.

Our light SUSY particles are wino, bino, and gravitino, and the mass difference is not so big.
The lightest particle is bino, and wino is heavier than bino. The mass difference is
$O(0.1) \times m_{3/2}$ GeV.
This might be one specific feature of
the gauge messenger scenario in $SU(5)$ GUT, as discussed in Ref. \cite{Bae}.

\acknowledgments
We are grateful to Hiroyuki Abe for useful discussions and comments.
This work  is supported by Grant-in-Aid for Scientific research from the Ministry of Education, Science, Sports, and Culture (MEXT), Japan, 
N0. 25400252 (T.K.) and No. 23104011 (Y.O.).


\appendix

\section{mass spectrums of the particles in the symmetry breaking sector}
\label{appendix1}
We investigate the mass matrices for the remnant fields in the symmetry breaking sector.
First, let us discuss $(Z, \widetilde Z)$ and $(\rho, \widetilde \rho)$ components.
We define $Z_{\pm}$ and $\rho_{\pm}$ as
\beq
Z_{\pm}= \frac{\widetilde{Z} \pm Z^{\dagger}}{\sqrt{2}},~\rho_{\pm}= \frac{\rho \pm \widetilde{\rho}^{\dagger}}{\sqrt{2}}.
\eeq
The fermion masses are given by
\begin{eqnarray}
{\cal L}_f&=& - \begin{pmatrix}  \ov{ \lambda_-} & \ov{ Z_+} &  \ov {\rho_+} \end{pmatrix} M^f_{+}  \begin{pmatrix} \lambda_- \\ Z_+  \\  \rho_+ \end{pmatrix}
 - \begin{pmatrix} \ov{\lambda_+}  &  \ov{Z_-} &  \ov{ \rho_-} \end{pmatrix} M^f_{-}   \begin{pmatrix}  \lambda_+ \\  Z_- \\   \rho_- \end{pmatrix},
 \end{eqnarray}
where the mass matrices $(M^f_{\pm})$ are
\beq
M^f_+=\begin{pmatrix}  0 & - g \Delta v & g v_{\chi} \\- g \Delta v & 0 & -h v_{\chi} \\ g v_{\chi}& -h v_{\chi} & -h \Delta v \end{pmatrix},~M^f_-=\begin{pmatrix} 0 & - g \Delta v &  g v_{\chi} \\  - g \Delta v &  0 &  h v_{\chi} \\   g v_{\chi} & h v_\chi & h \Delta v  \end{pmatrix},
\eeq
and $\lambda_{\pm}$ are the linear combinations of the gauginos $(X_{(+)})$ which are the suparpartners of $X_{\mu}$,
\beq
\lambda_{\pm}=\frac{X_{+} \pm \ov{X}}{\sqrt 2}.
\eeq

The masses for the bosonic superpartners are
\begin{eqnarray}
{\cal L}_B&=& - \begin{pmatrix}   Z^{\dagger}_+ &  \rho^{\dagger}_+ \end{pmatrix} M^2_{+}  \begin{pmatrix}  Z_+  \\  \rho_+ \end{pmatrix}
 - \begin{pmatrix}   Z^{\dagger}_- &  \rho^{\dagger}_- \end{pmatrix} M^2_{-}   \begin{pmatrix}   Z_- \\   \rho_- \end{pmatrix},
 \end{eqnarray}
where the mass matrices $(M^2_{\pm})$ are given by
\begin{eqnarray}
M^2_+&=&\begin{pmatrix} h^2 v_{\chi}^2  & -h^2 v_{\chi}\Delta v \\ -h^2 v_{\chi}\Delta v & h^2 (v_{\chi}^2+\Delta v^2)+F_X \end{pmatrix},  \\
M^2_-&=&\begin{pmatrix}  h^2 v_{\chi}^2+ g^2 \Delta v^2 & -(h^2+g^2)\Delta v v_\chi \\   -(h^2+g^2)\Delta v v_\chi & h^2 (v_{\chi}^2+\Delta v^2)+ g^2 v_\chi ^2 -F_X   \end{pmatrix}.
\end{eqnarray}
The F-term $F_X$ is $F_X=-h^2v^2_{\chi},$
so that $M^2_+$ includes the Goldstone mode.

The fermion masses for the other particles are also generated by the VEVs:
\begin{eqnarray}
{\cal L}_Y&=&  -\frac{1}{2} \begin{pmatrix} \widetilde W^A & Y^A  \end{pmatrix} M_Y \begin{pmatrix} \chi^A \\ \widetilde \chi^A\end{pmatrix}  -\frac{1}{2} \begin{pmatrix}   \chi^A & \widetilde \chi^A \end{pmatrix} M^T_Y \begin{pmatrix} \widetilde W^A \\ Y^A \end{pmatrix} +h.c.,
\end{eqnarray}
where $ \widetilde W$ is the superpartner of $W'$ and $M_{Y}$ are defined as
\begin{eqnarray}
M_Y&=&\left(
\begin{array}{cc}
-\frac{1}{\sqrt 2}M_{W'} &\frac{1}{\sqrt 2} M_{W'}   \\
-h v_\chi & -hv_\chi \\
\end{array}
\right). 
\end{eqnarray}
 The eigenvalues are
$M_{W'},~M_{W'},~\sqrt{2}h v_{\chi},~\sqrt{2}h v_{\chi}$ and 
the bosonic masses are given by the same mass spectrum.
The imaginary part of $\chi-\widetilde \chi$ corresponds to the Goldstone boson, and the real part
has the mass, $M_{W'}$, according to the D-term. The other masses, $\sqrt{2}h v_{\chi}$, correspond
to the ones of $\chi+\widetilde \chi$ and $Y$.

The singlet components $(Y_0, \chi_0, \widetilde{\chi}_0)$ of $\Hat Y$ and $( \Hat{\chi},\Hat{ \widetilde{\chi}})$ also get masses, according to the nonzero $v_{\chi}$. The fermionic mass matrix is
\begin{eqnarray}
{\cal L}_{Y_0}&=&  -\frac{1}{2} \begin{pmatrix} \widetilde {Z'} & Y_0  \end{pmatrix} M_{Y_0} \begin{pmatrix} \chi_0 \\ \widetilde {\chi_0}\end{pmatrix}  -\frac{1}{2} \begin{pmatrix}   \chi_0 & \widetilde{ \chi_0} \end{pmatrix} M^T_{Y_0} \begin{pmatrix} \widetilde {Z'} \\ Y_0 \end{pmatrix} +h.c.,
\end{eqnarray}
where  $M_{Y_0}$ are defined as
\begin{eqnarray}
M_{Y_0}&=&\left(
\begin{array}{cc}
-\frac{1}{\sqrt 2}M_{Z'} & \frac{1}{\sqrt 2}M_{Z'}   \\
-h v_\chi & -hv_\chi \\
\end{array}
\right). 
\end{eqnarray}
 The mass spectrums are given, relplacing $M_{W'}$ with $M_{Z'}$.


\section{Concrete parameter set}\label{appendix2}
The parameter sets which predict $m_h \approx 125$ GeV are  in Table \ref{table4}.
The Higgs mass is calculated by FeynHiggs \cite{Higgs3, FeynHiggs}.
 $m_{\widetilde{t}_{1,2}}$ are the stop masses in the mass eigenstate.
 $m^2_{\widetilde{Q}_L}$, $m^2_{\widetilde{d}_R}$, $m^2_{\widetilde{l}_L}$
and $m^2_{\widetilde{e}_R}$ are the soft SUSY breaking terms of the squarks
$(\widetilde{Q}_L, \widetilde{d}_R)$ and sleptons $(\widetilde{l}_L,\widetilde{e}_R)$.

\begin{table}[th]
\begin{center}
\begin{tabular}{|c|c|c|c|c|} \hline
& $N_{eff}=6$ & $N_{eff}=6$ & $N_{eff}=6.97$ & $N_{eff}=7.83$  \\  \hline
 $m_{3/2}$   & $588.84$ GeV & $741.31$ GeV   & $495.79$ GeV  & $245.02$ GeV \\  \hline
 $T_{\rho}$  & $2.39 \times 10^{11}$ GeV & $5.76 \times 10^{12}$ GeV& $5.66 \times 10^{12}$ GeV & $2.74 \times 10^{10}$ GeV  \\ \hline
  $T_{X}$ & $1.00 \times 10^{7}$ GeV & $1.00 \times 10^{7}$ GeV & $1.00 \times 10^{7}$ GeV  & $ 1.00\times 10^{7}$ GeV  \\ \hline
 $\tan \beta$ & $3.69$ & $3.93$ & $3.43$ & $4.04$   \\  \hline 
 \hline
 $m_h$  & $126.20$ GeV & $125.89$ GeV & $124.65$ GeV & $124.03$ GeV \\ \hline
  $m_{stop}$ & $3.05$ TeV & $3.61$ TeV & $2.93$ TeV &  $1.90$ TeV  \\  \hline
 $X_t$ & $ 3.43  \times$ $m_{stop}$ & $ 3.41  \times$ $m_{stop}$  & $ 2.98 \times$ $m_{stop}$ & $2.39 \times$ $m_{stop}$  \\  \hline
  $|\mu|$  & $3.72$ TeV & $4.38$ TeV& $3.27$ TeV & $1.93$ TeV \\ \hline
   $|B|$ & $4.21$ TeV & $4.72$ TeV & $3.22$ TeV & $1.97$ TeV \\ \hline
    $|B_{\rm{EW}}/B|$ & $0.92$  & $1.00$ & $0.91$ & $0.5$ \\ \hline
     \hline
 $|M_3|$&  $5.73 \times m_{3/2}$ &  $5.73 \times m_{3/2}$ & $6.85 \times m_{3/2}$   & $7.83 \times m_{3/2}$  \\ \hline 
  $|M_2|$   & $0.98 \times m_{3/2}$  & $0.98 \times m_{3/2}$& $0.98 \times m_{3/2}$  & $0.98 \times m_{3/2}$  \\ \hline 
   $|M_1|$ & $0.75 \times m_{3/2}$  & $0.75 \times m_{3/2}$ & $0.69 \times m_{3/2}$ & $0.64 \times m_{3/2}$  \\ \hline 
   $m_{\widetilde{t}_1}$ & $3.52$ TeV & $4.12$ TeV & $3.31$ TeV & $2.15$ TeV  \\ \hline 
  $m_{\widetilde{t}_2}$  & $2.62$ TeV & $3.17$ TeV& $2.57$ TeV & $1.65$ TeV  \\ \hline 
$m^2_{\widetilde{Q}_L}$& $11.72$ TeV$^2$ & $16.56$ TeV$^2$ & $10.36$ TeV$^2$ & $4.24$ TeV$^2$  \\ \hline
    $m^2_{\widetilde{d}_R}$ &  $15.97$ TeV$^2$ &  $22.52$ TeV$^2$ &  $13.60$ TeV$^2$ & $5.40$ TeV$^2$  \\ \hline 
          $m^2_{\widetilde{l}_L}$&  $0.78$ TeV$^2$ &  $0.93$ TeV$^2$ &  $0.44$ TeV$^2$ & $0.18$ TeV$^2$  \\ \hline
    $m^2_{\widetilde{e}_R}$&  $1.42$ TeV$^2$ &  $1.75$ TeV$^2$ &  $0.81$ TeV$^2$ & $0.31$ TeV$^2$  \\ \hline
 \end{tabular}
\caption{
\label{table4}%
{
SUSY mass spectrums and parameters with $\Lambda_{GUT}= 2 \times 10^{16}$ GeV.
Higgs mass is calculated by FeynHiggs \cite{Higgs3, FeynHiggs}.
}
}
\end{center}
\end{table}

\end{document}